\documentclass[aps,pre,floatfix,superscriptaddress,twocolumn,10pt]{revtex4-2}
\usepackage{color}
\usepackage{graphicx}
\usepackage{amsmath, amsthm, amssymb}
\usepackage[colorlinks,citecolor=red,urlcolor=blue]{hyperref}
\newcommand{\be}{\begin{equation}}
\newcommand{\ee}{\end{equation}}
\newcommand{\bea}{\begin{eqnarray}}
\newcommand{\eea}{\end{eqnarray}}
\usepackage{booktabs}

\begin{document}

\title{Finite-size scaling and edge effects in the Takayasu model of aggregation–diffusion with input}

\author{Rohan Banerjee Ravindran}
\email{rr694@snu.edu.in}
\affiliation{Department of Physics, Shiv Nadar Institution of Eminence, Gautam Buddha Nagar, Uttar Pradesh 201314,
India}
\author{R. Rajesh} 
\email{rrajesh@imsc.res.in}
\affiliation{The Institute of Mathematical Sciences, C.I.T. Campus, Taramani, Chennai 600113, India}
\affiliation{Homi Bhabha National Institute, Training School Complex, Anushakti Nagar, Mumbai 400094, India}

\date{\today}

\begin{abstract}
We analytically and numerically study the effect of finite spatial boundaries on the Takayasu model of diffusing and aggregating particles with steady monomer input in one dimension. Exact expressions are derived for the steady-state density profile, two-point correlation functions, and mean-squared density under both open and periodic boundary conditions. The single-site mass distribution exhibits a crossover from a bulk power law $P(m)\sim m^{-4/3}$ to an edge power law $P(m)\sim m^{-5/3}$, occurring near the boundaries or the condensate that forms in periodic systems. The equivalence between the two boundary conditions is shown to break down in the case of multipoint probability distributions near the edge.  The exact solution identifies a distinct boundary layer and shows that the edge anomaly arises when spatial mass currents, which scale as $\mathcal{O}(L)$,
dominate over the $\mathcal{O}(1)$ constant flux in mass space. We further generalize these results to mass-dependent diffusion.
\end{abstract}

\maketitle

\section{\label{sec:introduction}Introduction}

The large-scale dynamics of irreversible aggregation has been studied since the pioneering work of Smoluchowski more than a century ago~\cite{smoluchowski1917mathematical} (see also~\cite{aldous1999deterministic,leyvraz2003scaling,wattis2006introduction,handbook} for reviews). When the depletion of clusters through aggregation is balanced by a continual input of small clusters (monomers), the system can evolve into a nontrivial nonequilibrium steady state. Such aggregation-with-input processes arise in diverse physical and biological settings, including cloud microphysics~\cite{pruppacher1998microphysics}, surface growth~\cite{steyer1991growth}, planetary rings~\cite{brilliantov2015size,connaughton2018stationary}, atmospheric aerosols~\cite{hidy2013dynamics,drake1972topics}, soot agglomeration~\cite{friedlander2000smoke,sorensen2018light}, and the self-assembly of biomolecular condensates and cell aggregates~\cite{brangwynne2011active,lee2023size,arsenadze2024anomalous,liu2021scale,leggett2019motility}.

A paradigmatic model capturing this class of phenomena is the Takayasu model~\cite{takayasu1989apparent,takayasu1989steady,hayakawa1987irreversible}—also known as cluster–cluster aggregation with input—in which diffusing masses coalesce upon contact while being sustained by a constant injection of monomers. Beyond its physical relevance, the model maps exactly to several other systems, including river networks~\cite{scheidegger1967stochastic,dodds1999unified}, force propagation in granular packs~\cite{coppersmith1996model,rajesh2000exact}, the voter model~\cite{liggett2013stochastic}, and directed abelian sandpile models~\cite{dhar1989exactly,dhar1999abelian}. An important and fruitful analogy also exists with turbulence~\cite{hayakawa1987irreversible,connaughton2004stationary}: the mass distribution plays the role of the energy spectrum, aggregation acts as a nonlinear transfer term, and the steady state corresponds to a constant flux of mass through mass space. Within this analogy, the Takayasu model provides a minimal setting in which to study the breakdown of Kolmogorov scaling~\cite{connaughton2005breakdown,connaughton2006cluster} and to generalize the Kolmogorov $4/5$ law to constant-flux relations in driven dissipative systems~\cite{connaughton2007constant}.

In the infinite-system limit ($L \to \infty$ followed by $t \to \infty$), the model self-organizes into a critical steady state with a power-law mass distribution $P(m) \sim m^{-\tau_1}$. The exact solution in one dimension gives $\tau_1 = 4/3$~\cite{takayasu1989steady,huber1991scheidegger}. In general dimension $d<2$, exact and renormalization-group results yield $\tau_1 = (2d+2)/(d+2)$~\cite{rajesh2000exact,connaughton2005breakdown}, while for $d=2$ the upper-critical dimension, $P(m) \sim m^{-3/2} \sqrt{\ln m}$, and for $d>2$, the mean-field value $\tau_1 = 3/2$ holds. Higher-order joint distributions $P(m_1,\ldots,m_n)$ have the homogeneity exponent $\tau_n = n\tau_1 + [(2-d)/(2+d)]n(n-1)/2$ for $d<2$~\cite{connaughton2005breakdown}. The universality of these steady states depends on a locality criterion for the aggregation kernel $K(m_1,m_2)$: when interactions between vastly different mass scales dominate, universality breaks down~\cite{connaughton2004stationary}.

The role of finite boundaries, however, remains far less understood. Boundaries can be introduced either in mass space or in real space. A cutoff in mass space can destabilize the steady state, giving rise to oscillatory kinetics where mass is periodically transferred through the system~\cite{ball2012collective}. Finite spatial systems ($t \to \infty$ before $L \to \infty$), by contrast, reveal a different class of finite-size effects.  In one-dimensional systems with periodic boundaries, it was shown numerically that beyond a crossover time $t_c \sim L^2$, nearly all injected mass is absorbed by a single aggregate—the \emph{condensate}~\cite{negi2024condensate}. The mean total mass excluding this condensate reaches a steady value, while the condensate continues to grow linearly in time. The condensate imposes an effective boundary condition on the fluctuating mass field, reshaping its local environment. In the co-moving frame of the condensate, the single-site mass distribution near it ($x/L \lesssim 0.1$) follows a distinct power law, $P(m) \sim m^{-\tau_1}$ with $\tau_1 \approx 5/3$, in contrast to the bulk exponent $\tau_1 = 4/3$. Open systems also show signatures of finite-size. It was shown that the total mass exhibits strong intermittency due to successive build-ups and crashes~\cite{negi2024condensate}. Related dynamic condensates have also been reported in infinite systems during coarsening~\cite{das2024dynamic}, where the largest cluster carries a macroscopic fraction of the total mass and exhibits logarithmic corrections consistent with extremal-value statistics.

These observations raise key questions. Can the exponents $\tau_1 = 4/3$ and $\tau_1 = 5/3$, obtained numerically for finite periodic systems, be derived analytically from first principles? For open boundaries, are there corresponding exact  site-dependent exponents characterizing bulk and edge behavior? How are multipoint joint distributions modified by finite boundaries? 

In this paper, we present an exact calculation of the mean density and two-point correlations for the Takayasu model in one dimension under both open and periodic boundary conditions. This analysis yields exact bulk and edge exponents, and, based on controlled assumptions, allows us to infer higher-order scaling exponents $\tau_n$, which we verify using Monte Carlo simulations. These results are generalized to the case of mass-dependent diffusion. Finally, we discuss the implications for finite-size effects in higher dimensions.

The remainder of the paper is organized as follows. Section~\ref{sec:model} defines the model. The analytical calculations for the mean density and two-point correlations as well as the determination of the different exponents are presented for the open boundary conditions in Sec.~\ref{sec:obc} and periodic boundary conditions in Sec.~\ref{sec:pbc}. The results are generalized to the case when the diffusion constant is mass dependent in Sec.~\ref{sec:mass-dependent}. We conclude in Sec.~\ref{sec:conclusions} with a summary and discussion of  broader implications and open questions.

\section{\label{sec:model} Model}

We define the model in one dimension. Generalization to higher dimensions is straightforward. Consider a chain of lattice sites, each capable of holding a non-negative integer mass $m_i$. Initially, all sites are empty. The system evolves through two competing stochastic processes: uniform rate $J$ for each lattice site of injecting mass $1$, and random hopping to nearest-neighbor sites at rate $D$ ($D/2$ to each side). When a mass hops to an occupied site, the two masses coalesce, preserving their sum.

In this paper, we examine the model on finite chains with open or periodic boundary conditions. For open boundary conditions, we consider a chain of $L-1$ sites. Masses reaching either end of the chain are permanently lost from the system. Mass conservation is explicitly broken through both the continuous injection process and boundary dissipation, creating a non-equilibrium system with a spatially varying mass distribution. For periodic boundary conditions, we consider $L$ sites arranged in a ring geometry, where mass is strictly conserved apart from the external injection, and translation symmetry is maintained. In this case, a spatially dependent mass distribution arises only if the system spontaneously breaks translation invariance.

The choice of $L-1$ sites for open boundary conditions and $L$ sites for periodic boundary conditions ensures that once a condensate forms in the periodic case, the number of sites in the fluid phase is the same for both boundary conditions. In our calculations, we also take the continuum limit of lattice spacing $a \to 0$ and $L\to \infty$ such that $a L$ is finite. For notational convenience, we will denote this length ($a L$) also by $L$, as the context makes it clear what $L$ corresponds to.

\section{\label{sec:results}Results}

In this section, we systematically analyze the model for both open and periodic boundary conditions. We derive the steady-state density and two-point correlations functions which allows us to determine the site-dependent scaling exponents. Finally, we extend our analysis to multipoint correlation functions to probe higher-order correlations. 

The site dependent mass distributions are determined by assuming the scaling form:
\begin{equation}
P(m,L)=m^{-\tau_1}f\left(\frac{m}{L^{\phi}}\right),
\label{eq1}
\end{equation}
where $\tau_1$ and $\phi$ are site-dependent exponents and $f(x)\sim$ constant for $x\ll1$ and $f(x)\to 0$ for $x\gg 1$. Then $\langle m\rangle \sim L^{\phi(2-\tau_1)}$ and $\langle m^2 \rangle \sim L^{\phi(3-\tau_1)}$. Knowing $\langle m_i\rangle$ and $\langle m_i^2 \rangle$ gives two relationships for $\phi, \tau_1$, thus solving them fully.

To examine the effect of correlations, we will also define multipoint probability distributions $P(m_1, m_2, \ldots, m_n)$, when $m_1, m_2, \ldots, m_n$ are the masses on $n$ consecutive sites. We define the homogeneity exponent $\tau_n$ as
\be
P(h m_1, h m_2, \ldots, h m_n)= h^{-\tau_n} P(m_1, m_2, \ldots, m_n),
\label{eq2}
\ee
consistent with the power law $m^{-\tau_1}$ in Eq.~(\ref{eq1}) for the single site mass distribution. In simulations, it is more convenient to measure the complementary cumulative distribution function (CCDF)
\be
F_n(m) = \int_m^{\infty} dm_1 \ldots \int_m^{\infty} dm_n P(m_1, m_2, \ldots, m_n).
\label{eq3}
\ee
The CCDF will be assumed to have the scaling form 
\begin{equation}
F_n(m,L)=m^{-\zeta_n} g_n\left(\frac{m}{L^{\phi}}\right)
\label{eq4}
\end{equation}
where $g_n(x)\sim$ constant for $x\ll1$ and $g_n(x)\to 0$ for $x\gg 1$. Clearly
\be
\zeta_n=\tau_n-n.
\label{eq5}
\ee

\subsection{\label{sec:obc}Open boundary conditions}
\subsubsection{\label{sec:densityprofile}Density profile}
The evolution of the mass for sites $i=1,2,\ldots, L-1$ follows the stochastic equation
\begin{equation}
\begin{split}
&m_i(t+dt) -m_i(t)= \\
&m_{i-1} \eta^+_{i-1} + m_{i+1}\eta^-_{i+1} - m_i\eta^+_i - m_i\eta^-_i + I_i,
\end{split}
\label{eq6}
\end{equation}
where $m_0=m_L=0$, $\eta^{\pm}_i$ represents a binary hopping event where the $i$th mass either jumps to a neighboring site or remains in place, and $I_i$ denotes the injection process. These stochastic variables have the statistics
\begin{equation}
\langle\eta^{\pm}_i\rangle =\frac{Ddt}{2} , ~ \langle I_i\rangle = J dt,
~\langle(\eta^{\pm}_i)^2\rangle =\frac{Ddt}{2} , ~ \langle I_i^2\rangle = J dt.
\label{eq7}
\end{equation}
All other two point correlations are of the order $dt^2$ and therefore are set to zero.

Taking the  average of Eq.~(\ref{eq6}) and defining $\rho_i \equiv \langle m_i\rangle$ as the mean mass density yields
\begin{equation}
\frac{d\rho_i}{dt} = \frac{D}{2} \left(\rho_{i-1} + \rho_{i+1} - 2\rho_i\right) + J.
\label{eq8}
\end{equation}
In the steady state ($d\rho_i/dt = 0$), this reduces to the discrete Poisson equation:
\begin{equation}
\rho_{i-1} + \rho_{i+1} - 2\rho_i = -\frac{2J}{D}.
\label{eq9}
\end{equation}

\begin{figure}
\includegraphics[width=\columnwidth]{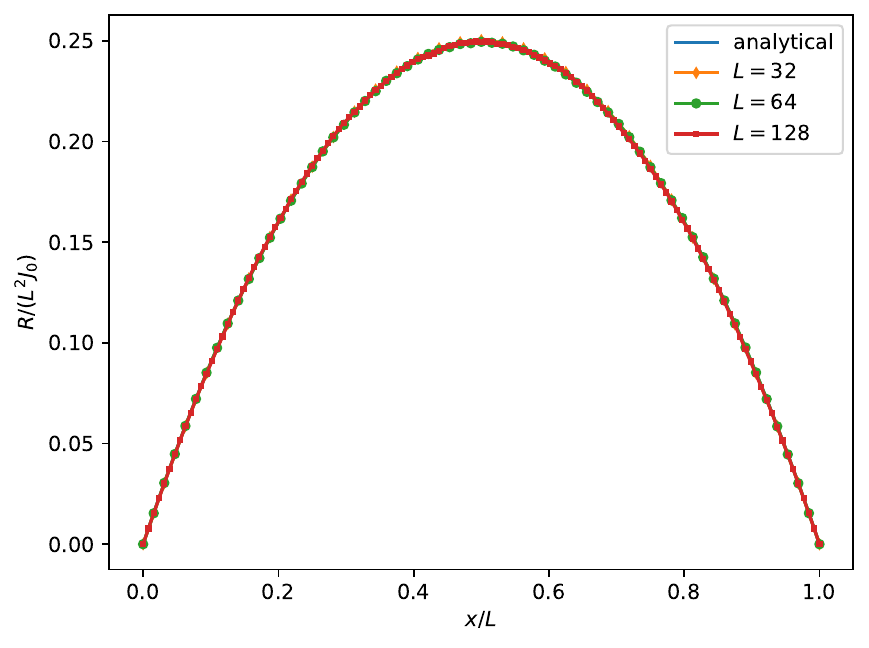}
\caption{The scaled density, $R(x)$, obtained from Monte Carlo simulations is compared with the exact result, Eq.~(\ref{eq12}). The data are for open boundary conditions.}
\label{fig:R}
\end{figure}

Taking the continuum limit where the lattice spacing $a\to0$ and $L \to \infty$ while maintaining constant system length $L a$ (also denoted as $L$), we define the spatial coordinate $x = ia$ and the continuous density profile $R(x) = a^{-1}\rho_{x/a}$. The diffusion constant and injection rate are renormalized as $\tilde{D}=Da^2$ and $\tilde{J}=J/a$ respectively. Substituting these into Eq.~(\ref{eq9}) yields:
\begin{equation}
R(x-a) + R(x+a) - 2R(x) = -\frac{2\tilde{J}}{\tilde{D}}a^2.
\label{eq10}
\end{equation}
Taylor expanding for small $a$, and then letting $a\to 0$ gives the
differential equation
\begin{equation}
\frac{d^2R}{dx^2} = -2J_0, \quad J_0 \equiv \frac{\tilde{J}}{\tilde{D}},
\label{eq11}
\end{equation}
where $J_0$ represents the ratio of injection to diffusion rates. We solve Eq.~(\ref{eq11}) subject to the open boundary conditions $R(0)=R({L})=0$ to obtain
\begin{equation}
R(x) = J_0x({L}-x)
\label{eq12}
\end{equation} 
as the unique solution, vanishing linearly at both boundaries as seen in Fig.~\ref{fig:R}.

\subsubsection{Two-point mass correlation}
The evolution of the two point mass-mass correlation for all $i,j$ can be derived by multiplying Eq.~(\ref{eq6}) for both $i$ and $j$, taking the ensemble average and applying the steady state condition. We separately consider the cases $i \neq j$ and $i=j$.

For sites separated by at least one lattice spacing ($|i-j|\geq 2$), the correlation function $\tilde{C}(i,j)\equiv\langle m_im_j\rangle$ satisfies:
\begin{equation}
-4\tilde{C}(i,j) + \tilde{C}(i \pm 1,j)  + \tilde{C}(i,j \pm1)  = \frac{-2J}{D}(\rho_i + \rho_j).
\label{eq13}
\end{equation}
Nearest-neighbor pairs ($j=i+1$) follow a modified version:
\begin{equation}
\begin{split}
-4\tilde{C}(i,i+1)+\tilde{C}(i-1,i+1)&+\tilde{C}(i,i+2)\\
&=\frac{-2 J}{D}(\rho_i+\rho_{i+1}).
\label{eq14}
\end{split}
\end{equation}
Notice that setting $\tilde{C}(i,i)\equiv 0$ for all $i$ makes Eq.~(\ref{eq13}) reduce exactly to the nearest-neighbor case in Eq.~(\ref{eq14}) when $j=i+1$. The symmetry $\tilde{C}(i,j)=\tilde{C}(j,i)$ means solving for $j>i$ suffices.

The diagonal case ($i=j$) introduces second moments $\tilde{E}_i\equiv\langle m_i^2\rangle$ through:
\begin{equation}
\begin{split}
-2\tilde{E}_i+\tilde{E}_{i+1}+&\tilde{E}_{i-1}+2\tilde{C}(i-1,i)\\
&+2\tilde{C}(i,i+1)
=\frac{-2 J}{D}(2\rho_i+1)
\label{eq15}
\end{split}
\end{equation}
This coupling demonstrates how prior solution of the two-point correlations $\tilde{C}(i,j)$ is required for determining the second moments $\tilde{E}_i$. 

Taking the continuum limit,
\begin{equation}
C(x,y)=\frac{1}{a^2} \tilde{C}\left(\frac{x}{a},\frac{y}{a}\right), \ \ \ \ \ \ E(x)=\frac{1}{a} \tilde{E}_{\frac{x}{a}}.
\label{eq16}
\end{equation}
and substituting into Eqs.~(\ref{eq13}) and (\ref{eq15}) and taking the limit $a\to 0$, we obtain
\begin{equation}
\frac{\partial^2 C(x,y)}{\partial x^2}+\frac{\partial^2 C(x,y)}{\partial y^2}=-2J_0[R(x)+R(y)]
\label{eq17}
\end{equation}
where $J_0$ is as defined in Eq.~(\ref{eq11}). The boundary conditions are $C(i,i)=C(i,{L})=C(0,j)=0$. Also,
\begin{equation}
\frac{\partial^2 E(x)}{\partial x^2}=-4\frac{\partial C(x,y)}{\partial y} \left.\right\vert_{y=x^+}-2J_0,
\label{eq18}
\end{equation}
with the boundary conditions $E(0)=E({L})=0$.

We begin solving for $C(x,y)$ with the boundary conditions defined on the  triangular domain  $0< x <y < L$. We closely follow the techniques in Ref.~\cite{sachdeva2014analytical} where similar equations were solved for the problem of aggregation with boundary input and exit. It is convenient to extend the region of validity  to the full square and consider the function $W(x,y)$ satisfying the equation:
\begin{equation}
\nabla^2 W(x,y)=-2J_0[R(x)+R(y)]\Theta(y-x),
\label{eq19}
\end{equation}
and therefore,
\begin{equation}
\begin{split}
\nabla^2[W(x,y)&-W(y,x)]\\
&=-2J_0[R(x)+R(y)](2\Theta(y-x)-1).
\end{split}
\label{eq20}
\end{equation}
Note that in the upper triangle where $(y>x)$, $W(x,y)-W(y,x)$ follows the exact same differential equation and boundary conditions as $C(x,y)$ which is what we need. Therefore to obtain 
\begin{equation}
C(x,y)\equiv W(x,y)-W(y,x),
\label{eq21}
\end{equation}
solving for $W(x,y)$ is sufficient.

To solve for $W(x,y)$, we start with the Fourier expansion for these type of boundary conditions,
\begin{equation}
\centering
W(x,y)=\sum_{i=1}^{\infty}\sum_{j=1}^{\infty}a_{ij}\psi_{ij}(x,y),
\label{eq22}
\end{equation}
where
\begin{equation}
\psi_{ij}(x,y)=\frac{2}{L} \sin{\frac{i\pi x}{L}} \sin{\frac{j\pi y}{L}},
\label{eq23}
\end{equation}
follow the orthonormality conditions,
\begin{equation}
\int_0^L\int_0^L \psi_{ij}(x,y)\psi_{i^{\prime}j^{\prime}}(x,y)dx dy=\delta_{i i^{\prime}}\delta_{jj^{\prime}}.
\label{eq24}
\end{equation}
Substituting into Eq.~(\ref{eq19}), multiplying both sides by $\psi_{ij}(x,y)$ and integrating over the square yields,
\begin{equation}
a_{ij}=\frac{2J_0^2 L^2}{\pi^2(i^2+j^2)}f_{ij},
\label{eq25}
\end{equation}
where
\begin{equation}
f_{ij}=\int^L_0 dx \int^L_x dy \  \psi_{ij}(x,y)[x(L-x)+y(L-y)],
\label{eq26}
\end{equation}
which can be evaluated by direct integration and are tabulated in Table~\ref{table1}.
\begin{table}
\caption{Values of $f_{ij}$ and $h_{ij}$ for different parities of $i,j$ \label{tab:parity}}
\begin{ruledtabular}
\begin{tabular}{cccc}
$i$ & $j$ & $f_{ij}$ & $h_{ij}$ \\ 
\midrule
odd & odd & $\frac{8L^3(i^2+j^2)}{i^3j^3\pi^4}$ & $0$ \\
odd & even & $\frac{8L^3j(9i^4-2i^2j^2+j^4)}{i^3(i^2-j^2)^3\pi^4}$ & $\frac{16L^3j(9i^4-2i^2j^2+j^4)}{i^3(i^2-j^2)^3\pi^4}$ \\
even & odd &  $\frac{8L^3i(9j^4-2i^2j^2+i^4)}{j^3(i^2-j^2)^3\pi^4}$ & $\frac{16L^3i(9j^4-2i^2j^2+i^4)}{j^3(i^2-j^2)^3\pi^4}$ \\
even & even & $0$ & $0$ 
\end{tabular}
\end{ruledtabular}
\label{table1}
\end{table}

Substituting $a_{ij}$ back into  Eq.~(\ref{eq22}), we obtain $C(x,y)\equiv W(x,y)-W(y,x)$ to be
\begin{equation}
C(x,y)=\sum_{ij}a_{ij}\psi_{ij}(x,y)-\sum_{ij}a_{ij}\psi_{ij}(y,x).
\label{eq27}
\end{equation}
But $\psi_{ij}(y,x)$ is the same as $\psi_{ji}(x,y)$. Substituting,
\bea
C(x,y)&=&\sum_{ij}(a_{ij}-a_{ji})\psi_{ij}(x,y) \label{eq28} \\
&=& \frac{2J_0^2L^2}{\pi^2(i^2+j^2)}h_{ij} \psi_{ij}(x,y)
\label{eq29}
\eea
where $h_{ij}=f_{ij}-f_{ji}$ is tabulated in Table~\ref{table1}.
This gives us the final form of $C(x,y)$:
\bea
C(x,y)&=&\frac{64 J_0^2 L^4}{\pi^6}
\!\!\sum_{\substack{i\,\mathrm{odd}\\ j\,\mathrm{even}}}
\!\frac{i^8 - 2 i^6 j^2 + 18 i^4 j^4 - 2 i^2 j^6 + j^8}
     {i^3 j^3 (i^2 - j^2)^3 (i^2 + j^2)} \nonumber \\
&&\sin \left(\frac{i\pi x}{L}\right)
\sin \left(\frac{j\pi y}{L}\right).
\label{eq30}
\eea
Substituting into Eq.~(\ref{eq18}), integrating twice, and applying the boundary conditions $E(0)=E(L)=0$ yields:
\begin{widetext}
\begin{equation}
E(x)=\frac{128J_0^2L^5}{\pi^7} \sum_{i\text{ odd}}\sum_{j\text{ even}} 
\frac{i^8-2 i^6 j^2+18 i^4 j^4-2 i^2 j^6+j^8}{j^2 \left(i^2+j^2\right) \left(i^3-i j^2\right)^3}    \left[\frac{\sin{\frac{(i-j)\pi x}{L}}}{(i-j)^2} +\frac{\sin{\frac{(i+j)\pi x}{L}}}{(i+j)^2}\right]
+J_0 x (L-x).
\label{eq31}
\end{equation}
\end{widetext}

\begin{figure}
\includegraphics[width=\columnwidth]{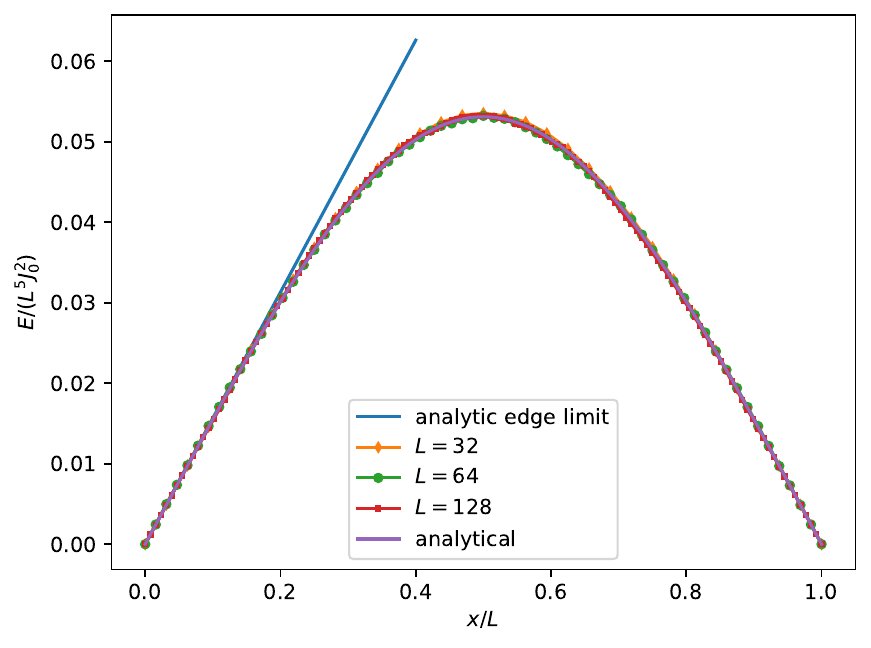}
\caption{The scaled second moment of density, $E(x)$, obtained from Monte Carlo simulations is compared with the exact result, Eq.~(\ref{eq31}). The solid line has slope $g$ as given in Eq.~(\ref{eq36}). The data are for open boundary conditions.
}
\label{fig:E}
\end{figure}

The results from Monte Carlo simulations are in excellent agreement with the analytical result, as can be seen from Fig.~\ref{fig:E}, confirming the correctness of Eq.~(\ref{eq31}). 

\subsubsection{\label{sec:exponents}Site-dependent exponents}

We now examine the scaling of  $E(x)$, distinguishing bulk ($x/L$ constant) and edge ($x$ fixed) regimes.

\paragraph{\label{sec:bulk}Bulk scaling:}

In the bulk limit $x, L \to \infty$, $x/L$  fixed:
\begin{equation}
E(x) = \frac{128J_0^2 L^5}{\pi^7} f\left(\frac{x}{L} \right),
\label{eq32}
\end{equation}
where the scaling function $f(x/L)$ contains the summations in Eq.~(\ref{eq31}).  This combined with $\langle m \rangle \sim L^2$ from the density profile [see Eq.~(\ref{eq12})], constrains the exponents in Eq.~(\ref{eq1}). Matching moments yields $\phi(2-\tau_1)=2$ and $\phi(3-\tau_1)=5$, determining $\tau_1$ and $\phi$ in the bulk to be:
\begin{equation}
\tau_{1, obc}^{B}=\frac{4}{3},  \ \ \ \ \phi_{obc}^{B}=3,
\label{eq33}
\end{equation}
where $B$ stands for bulk.  The value of $\tau_1$ is identical to that of the infinite system, confirming that bulk behavior is universal.

\begin{figure}
\includegraphics[width=\columnwidth]{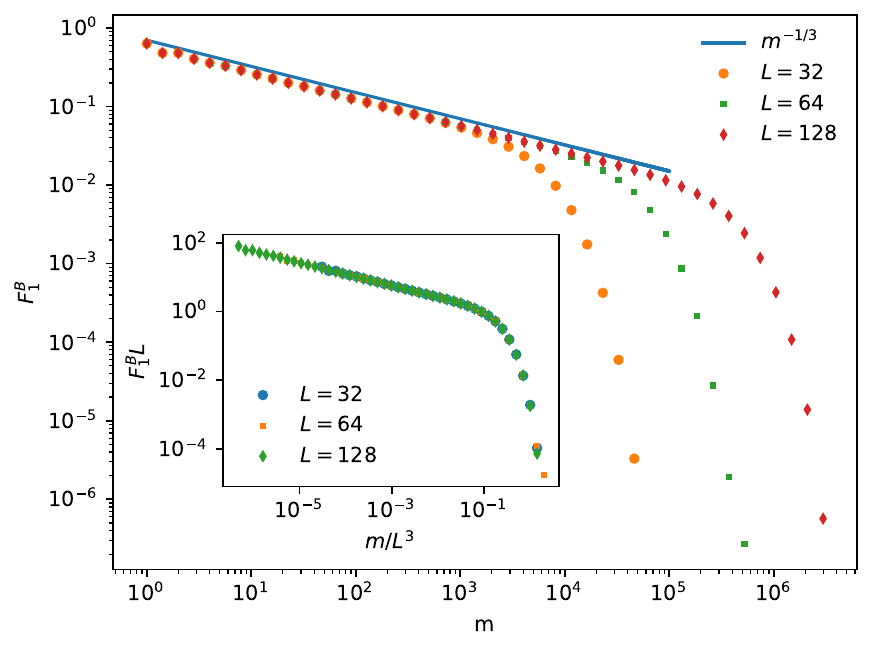}
\caption{The CCDF for the single site mass distribution, $F_1^B$, at the middle site $L/2$ for different system sizes. The data are consistent with the power law $m^{-1/3}$. Inset: The data for different $L$ collapse onto one curve when scaled with the exponents in Eq.~(\ref{eq33}). The data are for open boundary conditions.}
\label{fig:bulk-F1}
\end{figure}

The derivation of the exponents in Eq.~(\ref{eq33})  relies on the assumed finite-size scaling form of Eq.~(\ref{eq1}) and the power-law behavior of the scaling function. We confirm the correctness of the result with Monte Carlo simulations, as shown in Fig.~\ref{fig:bulk-F1}.  The data for the CCDF of the middle site $L/2$ for different system sizes collapse onto one curve when scaled with the exponents given in Eq.~(\ref{eq33}), with the scaling function having a power-law exponent $\tau_1-1=1/3$.

We now consider the multipoint probability distributions as defined in Eqs.~(\ref{eq2})--(\ref{eq4}). As for the single site distribution, we expect the bulk scaling for the multipoint probability distributions to be the same as for the infinite system. For the infinite, homogeneous system, it was shown, using renormalization group calculations~\cite{connaughton2005breakdown,connaughton2006cluster}, that in one dimension $F_n$ has the homogeneity exponent
\be
\zeta^B_n= \frac{n}{3} + \frac{n (n-1)}{6}, \quad n=1, 2, \ldots
\label{eq34}
\ee

\begin{figure}
\includegraphics[width=\columnwidth]{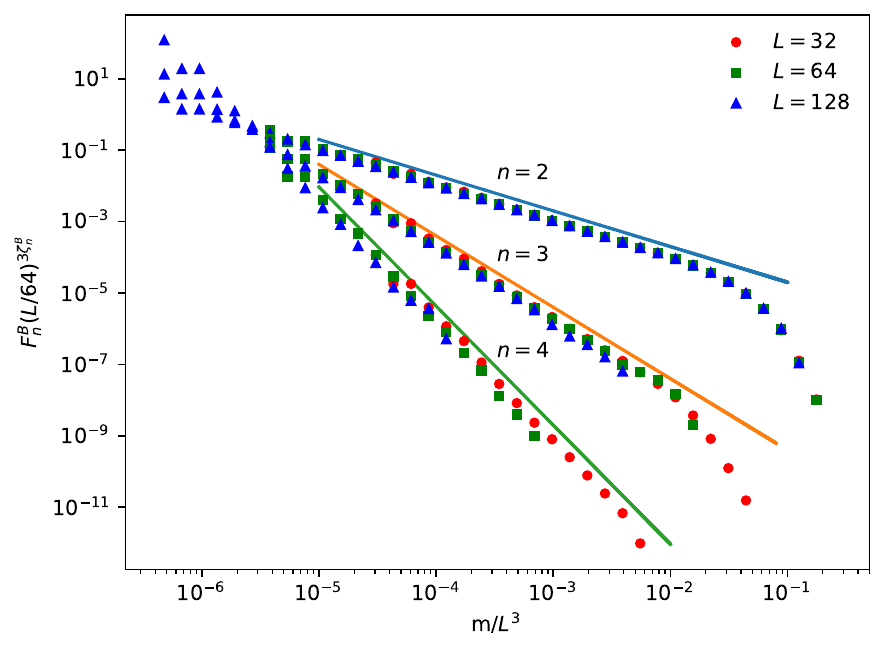}
\caption{The CCDF for multipoint probabilities, $F_n^B$ at the bulk middle site $L/2$ for different system sizes collapse onto one curve when scaled with the exponents in Eq.~(\ref{eq34}) for $n=2, 3, 4$. The solid lines correspond to power-laws with exponents $\zeta_n^B$. The data are for open boundary conditions.}
\label{fig:bulk-multi}
\end{figure}

We measure numerically $F_2$, $F_3$, and $F_4$ for the sites centered around the bulk middle site $L/2$. The data for different $L$ collapse onto a single curve when scaled with $\zeta^B_n$ as in Eq.~(\ref{eq34}) for $n=2, 3, 4$ (see Fig.~\ref{fig:bulk-multi}). We conclude that the scaling for the infinite system continues to hold for multipoint probability distributions in the bulk.

\paragraph{\label{sec:edge}Edge scaling:}

In the edge limit, $L\to \infty$ keeping $x$ fixed, the second moment of density scales as
\begin{equation}
E(x) = J_0^2 L^4 x g,
\label{eq35}
\end{equation}
where
\begin{equation}
g \equiv \sum_{i\text{ odd}}\sum_{j\text{ even}} \frac{512 j^2(9i^4-2i^2j^2+j^4)}{\pi^6 i^2(i^2-j^2)^4(i^2+j^2)}=0.156531\ldots.
\label{eq36}
\end{equation}

The finite value of $g$ establishes the $L^4$ scaling at the edge of the system for the second moment which contrasts with the bulk $L^5$ behavior. This, combined with $\langle m \rangle \sim L$ from the density profile [see Eq.~(\ref{eq12})], constrains the exponents in Eq.~(\ref{eq1}). Matching moments yields $\phi(2-\tau_1)=1$ and $\phi(3-\tau_1)=4$, determining $\tau_1$ and $\phi$ at the edge to be:
\begin{equation}
\tau_{1, obc}^{E}=\frac{5}{3},  \ \ \ \ \phi_{obc}^{E}=3,
\label{eq37}
\end{equation}
where $E$ stands for edge.

\begin{figure}
\includegraphics[width=\columnwidth]{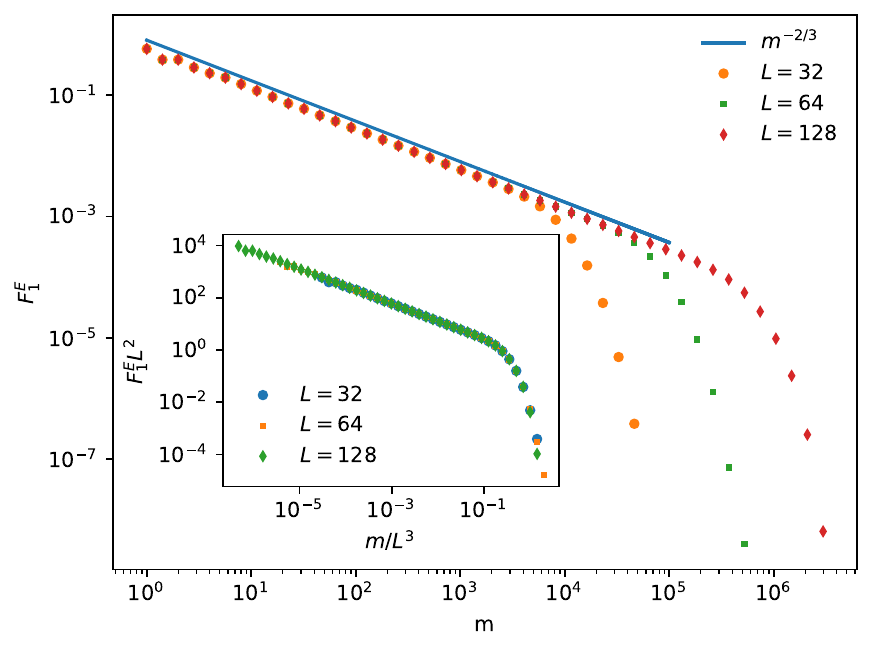}
\caption{The CCDF for the single site mass distribution, $F_1^E$ at the edge site $1$ for different system sizes. The data are consistent with the power law $m^{-2/3}$. Inset: The data for different $L$ collapse onto one curve when scaled with the exponents in Eq.~(\ref{eq37}). The data are for open boundary conditions.}
\label{fig:bulk-F2}
\end{figure}

We note that the value of $\tau_1$ is different from the bulk. We confirm the correctness of the result with Monte Carlo simulations, as shown in Fig.~\ref{fig:bulk-F2}. For the mass distribution of edge site $1$, the data for different system sizes collapse onto one curve when scaled with the exponents given in Eq.~(\ref{eq37}), with the scaling function having a power-law exponent $\tau_1-1=2/3$.

\begin{figure}
\includegraphics[width=\columnwidth]{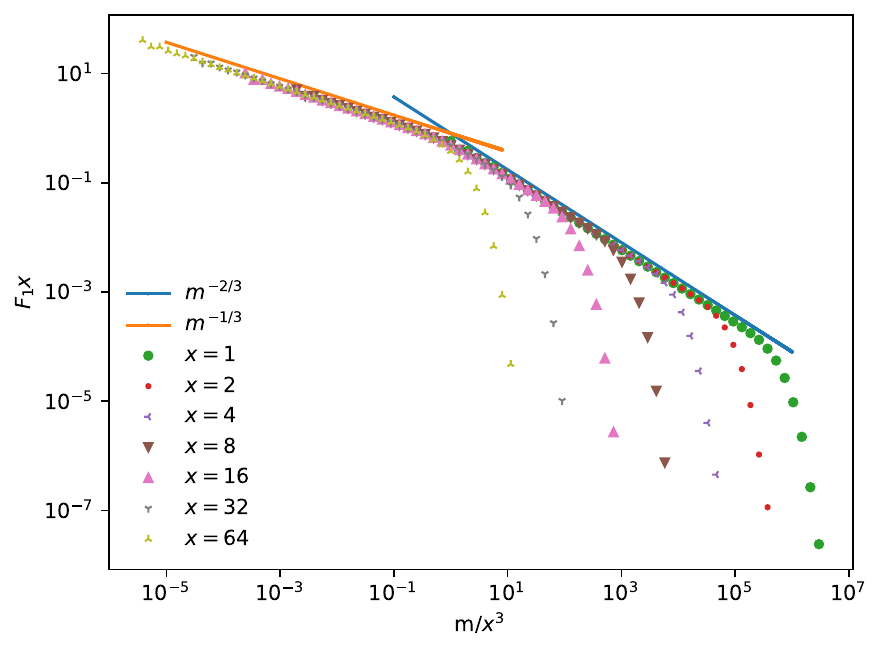}
\caption{The CCDF for the single site mass distribution, $F_1(m, x)$ for different $x$ collapse onto one curve when scaled as in Eq.~(\ref{eq38}). The straight lines are power laws describing the scaling function as argued in the text. The data are for $L=128$ with open boundary conditions.}
\label{fig:spacedependent}
\end{figure}

Given the different scaling in the bulk and edge, we now quantify the crossover from the bulk to edge scaling. Consider the site-dependent single site mass distribution $F_1(m,x)$. Since, $x$ and $L$ have the same dimension, we expect $F_1(m,x)$ to have the scaling form
\be
F_1(m,x) = \frac{1}{x^{\phi \zeta_1^B}} h\left(\frac{m}{x^\phi} \right).
\label{eq38}
\ee
In the bulk, i.e., when the argument of $h$ is small, we expect no dependence on $x$, implying that $h(y)\sim y^{-\zeta_1^B}$ for $y\ll 1$. When $y\gg 1$, corresponding to edge sites, we expect $h(y)\sim y^{-\zeta_1^B-1/\phi}$, so that the dependence of $F_1(m,x)$ on $x$ is linear. This is required so that the spatial mass current, proportional to the first derivative, is independent of $x$. Note that we consider fixed $L$.  $F_1(m,x)$ for different $x$ are shown in Fig.~\ref{fig:spacedependent}. When scaled as in Eq.~(\ref{eq38}), the data for different $x$ collapse onto one curve. The data for small and large argument are consistent with the expected asymptotic behavior  for $h(y)$.

We now examine the multipoint probability distributions for sites near the edge, as defined in Eqs.~(\ref{eq2})--(\ref{eq4}). In this regime, there is no direct analogue of the infinite-system calculation, nor an exact evaluation of the multipoint correlations. We therefore adopt a phenomenological approach.

Let $P_{i,i+1,\ldots,i+n-1}(m_1,m_2,\ldots,m_n)$ denote the $n$-point probability distribution, where the subscripts specify the lattice sites. The BBGKY hierarchy couples the $n$-point and $(n+1)$-point distributions. These equations have identical structure for both bulk and edge sites. However, the single-site distribution differs near the edge due to the presence of a finite mass current.  

Since the underlying hierarchy is unchanged, we assume that the relative scaling among the multipoint distributions remains the same in the bulk and at the edge, and that any difference arises solely from the modified single-site statistics. To incorporate this effect, we postulate that the edge distributions can be obtained from the bulk ones through a simple change of variables. This approach is motivated by Ref.~\cite{connaughton2007constant1}, which demonstrated that even without a constant mass-space flux, multipoint correlations can be related by a variable transformation if binary aggregation dominates.

In the present context, we conjecture that a transformation of the form $m \to m^y$ maps bulk to edge behavior. Using the exact single-site exponents, we find $y = 2$, leading to
\be
\zeta^E_{n, obc} = 2\,\zeta^B_n.
\label{eq39}
\ee

\begin{figure}
\includegraphics[width=\columnwidth]{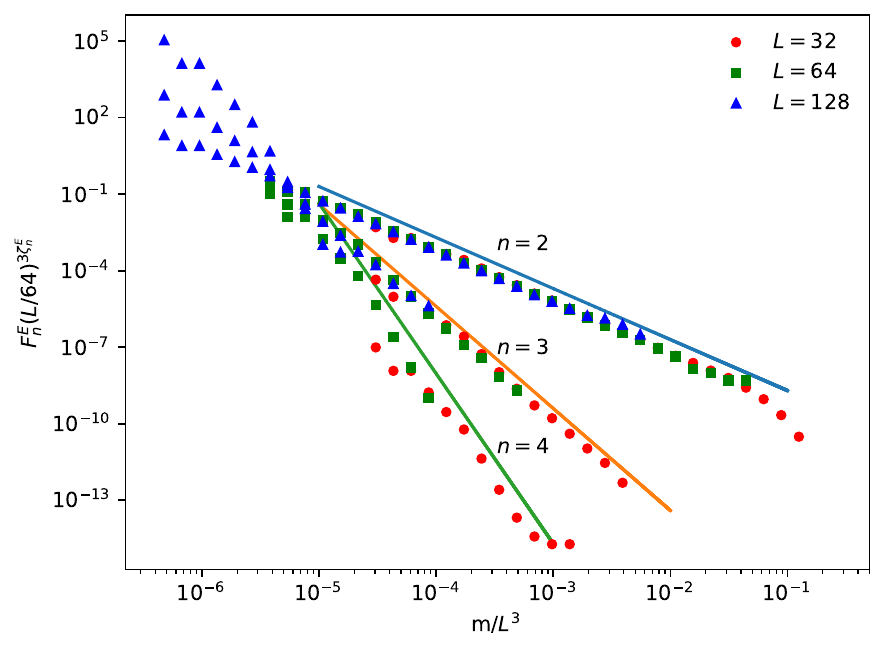}
\caption{The CCDFs for the multipoint probabilities, $F_n^E$, near the edge site ($x=1$) for different system sizes. The data collapse onto a single curve when scaled with the exponents in Eq.~(\ref{eq39}) for $n=2,3,4$. The solid lines indicate power laws with exponents $\zeta_n^E$. The data are for open boundary conditions.}
\label{fig:edge-multi}
\end{figure}

We numerically evaluate $F_2$, $F_3$, and $F_4$ for sites near the edge ($x=1$). As shown in Fig.~\ref{fig:edge-multi}, data for different system sizes collapse onto a single curve when scaled with $\zeta^E_{n, obc}$ as given by Eq.~(\ref{eq39}) for $n=2,3,4$. While measurements for $n=4$ are noisier due to the larger value of $\zeta^E_{4, obc}$, the available data exhibit a clear collapse over multiple decades, supporting the validity of Eq.~(\ref{eq39}). 

\subsection{\label{sec:pbc}Periodic boundary conditions}

Under periodic boundary conditions,  total mass is conserved apart from the continuous external injection. After time $t_C\sim L^2$, a condensate forms~\cite{negi2024condensate}. Without loss of generality we take site $0$ to be the location of the condensate and aim to calculate the mean density and mass correlations with respect to the condensate.
\subsubsection{\label{sec:densityprofile-pbc}Density profile}
The evolution of the mass for sites $i=1,2,\ldots, L-1$ follows the stochastic equation
\begin{equation}
\begin{split}
&m_i(t+dt) -m_i(t)= I_i+m_{i-1} \eta^+_{i-1} + m_{i+1}\eta^-_{i+1}- \\
&-m_i (\eta^+_i+\eta^-_i)+(m_{i+1}-m_i)\eta^-_C+(m_{i-1}-m_i)\eta^+_C,
\label{eq40}
\end{split}
\end{equation}
where $m_0=m_L=0$, $\eta^{\pm}_i$ and $I_i$ are defined under Eq.~(\ref{eq6}) and have the statistics Eq.~(\ref{eq7}). Here, $\eta^{\pm}_C$ represents hopping events where the condensate itself jumps, affecting all other masses due to re-labeling of sites. It also follows the same statistics as mentioned in Eq.~(\ref{eq7}). Taking the average of Eq.~(\ref{eq40}) yields
\begin{equation}
\frac{d\rho_i}{dt} = D \left(\rho_{i-1} + \rho_{i+1} - 2\rho_i\right) + J.
\label{eq41}
\end{equation}

\begin{figure}
\includegraphics[width=\columnwidth]{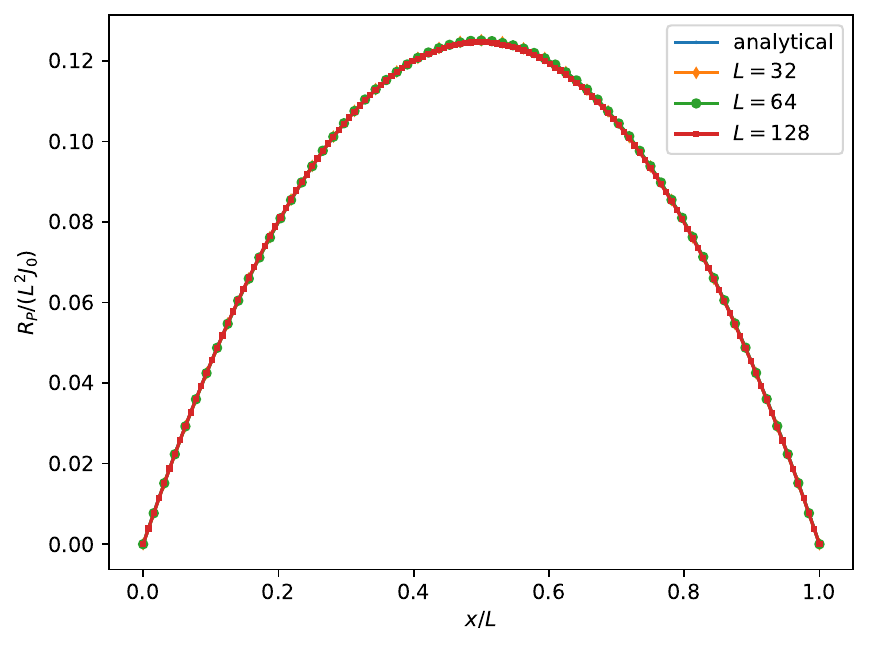}
\caption{The scaled density, $R_P(x)$, obtained from Monte Carlo simulations is compared with the exact result, Eq.~(\ref{eq43}). The data are for periodic boundary conditions. }
\label{fig:R_P}
\end{figure}

Taking the continuum limit just as we did under Eq.~(\ref{eq9}) and substituting into Eq.~(\ref{eq41}) yields:
\begin{equation}
\frac{d^2R_P}{dx^2} = -J_0
\label{eq42}
\end{equation}
where the subscript $P$ denotes the fact that we are working with periodic boundary conditions and $J_0$ is defined in Eq.~(\ref{eq11}). Solving Eq.~(\ref{eq42}) subject to $R_P(0)=R_P(L)=0$ gives a unique solution 
\begin{equation}
R_P(x)=\frac{J_0}{2}x(L-x),
\label{eq43}
\end{equation}
matching the result obtained in Ref.~\cite{negi2024condensate}.
$R_P(x)$ vanishes linearly at $x=0, L$, as seen in Fig.~\ref{fig:R_P}. We note that this result is identical to Eq.~(\ref{eq12}) obtained for open boundary conditions, apart from a multiplicative factor of two, arising from a renormalization of the diffusion constant due to the hopping of the condensate.

\subsection{Two-point mass correlation}

The evolution equation for the two-point mass–mass correlation can be obtained by multiplying Eq.~(\ref{eq40}) for both $i$ and $j$, taking the ensemble average, and imposing the steady-state condition. We consider separately the cases $i \neq j$ and $i = j$.

For sites separated by at least one lattice spacing ($|i-j| \geq 2$), the correlation function $\tilde{C}(i,j) \equiv \langle m_i m_j \rangle$ satisfies
\begin{equation}
\begin{split}
&-6\tilde{C}(i,j) + \tilde{C}(i \pm 1,j)  + \tilde{C}(i,j \pm 1) + \tilde{C}(i+1,j+1) \\
&  + \tilde{C}(i-1,j-1)
= -2\frac{J}{D}\,(\rho_i + \rho_j).
\end{split}
\label{eq44}
\end{equation}
Compared to the corresponding equation for open boundary conditions [see Eq.~(\ref{eq13})], there are two additional terms which are due to the coordinated movement of all the particles corresponding to the hopping of the aggregate.

For nearest-neighbor pairs ($j=i+1$), a modified relation holds:
\begin{equation}
\begin{split}
&-6\tilde{C}(i,i+1) + \tilde{C}(i-1,i+1) + \tilde{C}(i,i+2) \\
&+ \tilde{C}(i+1,i+2) + \tilde{C}(i-1,i)
= \frac{-2 J}{D}\,(\rho_i + \rho_{i+1}).
\end{split}
\label{eq45}
\end{equation}
Setting $\tilde{C}(i,i) \equiv 0$ for all $i$ makes Eq.~(\ref{eq44}) reduce exactly to Eq.~(\ref{eq45}) when $j = i + 1$. The symmetry $\tilde{C}(i,j) = \tilde{C}(j,i)$ implies that solving for $j>i$ is sufficient.

For the diagonal case ($i = j$), we introduce the second moments $\tilde{E}_i \equiv \langle m_i^2 \rangle$, which satisfy
\begin{equation}
\begin{split}
&-2\tilde{E}_i + \tilde{E}_{i+1} + \tilde{E}_{i-1}
+ \tilde{C}(i-1,i) + \tilde{C}(i,i+1)\\
&= \frac{-J}{D}(2\rho_i + 1).
\end{split}
\label{eq46}
\end{equation}
Thus, to determine $\tilde{E}_i$, we must first solve  two-point correlations $\tilde{C}(i,j)$.

Taking the continuum limit, we define
\begin{equation}
C_P(x,y) = \frac{1}{a^2}\,\tilde{C}\!\left(\frac{x}{a}, \frac{y}{a}\right),
\quad
E_P(x) = \frac{1}{a}\,\tilde{E}_{x/a},
\label{eq47}
\end{equation}
and substituting into Eqs.~(\ref{eq44}) and (\ref{eq46}) while taking $a \to 0$, we obtain
\begin{equation}
\nabla^2 C_P(x,y) + \frac{\partial^2 C_P(x,y)}{\partial y\,\partial x}
= -J_0 [R_P(x) + R_P(y)],
\label{eq48}
\end{equation}
where $J_0$ is defined in Eq.~(\ref{eq11}). The boundary conditions are
$C_P(x,x) = C_P(x,{L}) = C_P(0,y) = 0$.
Furthermore,
\begin{equation}
\frac{d^2 E_P(x)}{dx^2}
= -2\frac{\partial C_P(x,y)}{\partial y}\bigg|_{y=x^+}
- J_0,
\label{eq49}
\end{equation}
with $E_P(0) = E_P(L) = 0$.

To convert the regime of interest in Eq.~(\ref{eq48}) from the upper triangle to the full square, we follow  the same arguments used for the open boundary conditions [see Eqs.~(\ref{eq19})–(\ref{eq21})], and we solve the differential equation
\begin{equation}
\begin{split}
&\nabla^2 W_P(x,y)
+ \frac{\partial^2 W_P(x,y)}{\partial y\,\partial x}\\
&= -J_0 [R_P(x) + R_P(y)] \, \Theta(y - x).
\end{split}
\label{eq50}
\end{equation}
$C_P(x,y)$ is then given by
\begin{equation}
C_P(x,y) \equiv W_P(x,y) - W_P(y,x).
\label{eq51}
\end{equation}
To obtain $W_P(x,y)$, we expand it in the Fourier basis, as in the open boundary case [Eq.~(\ref{eq22})]:
\begin{equation}
W_P(x,y)
= \sum_{i=1}^{\infty}\sum_{j=1}^{\infty} a_{ij}\, \psi_{ij}(x,y),
\label{eq52}
\end{equation}
where $\psi_{ij}(x,y)$ is defined in Eq.~(\ref{eq23}).

In contrast to the open boundary condition case, the presence of the mixed derivative term in Eq.~(\ref{eq52}) makes determining the coefficients $a_{ij}$ nontrivial. Multiplying Eq.~(\ref{eq52}) by $\psi_{mn}(x,y)$ and integrating gives
\bea
&& \frac{a_{mn}}{mn}
- \sum_{m'} \sum_{n'}
\frac{4 a_{m'n'} m'n' [1 - (-1)^{m+m'}][1 - (-1)^{n+n'}]}
{\pi^2 (m^2 - m'^2)(n^2 - n'^2)(m^2 + n^2)}
\nonumber\\
&&= \frac{J_0^2 L^2}{2\pi^2 (m^2 + n^2) mn} f_{ij},
\label{eq53}
\eea
where $f_{ij}$ is listed in Table~\ref{table1}.

Defining the antisymmetric combination 
\be
c_{mn} =  \frac {\pi^6 (a_{mn} - a_{nm})}{8 J_0^2 L^5}.
\label{eq54}
\ee
we obtain
\bea
&& \frac{c_{mn}}{mn}
- \sum_{m'} \sum_{n'}
\frac{4 c_{m'n'} m'n' [1 - (-1)^{m+m'}][1 - (-1)^{n+n'}]}
{\pi^2 (m^2 - m'^2)(n^2 - n'^2)(m^2 + n^2)}
\nonumber\\
&&= \frac{ \pi^4 h_{ij}}{16 L^3 (m^2 + n^2) mn},
\label{eq55}
\eea
with $h_{ij}$ also given in Table~\ref{table1}.

The equations for $c_{mn}$ corresponding to odd–odd and even–even indices decouple from those with mixed parity. Since $h_{ij}=0$ whenever $i$ and $j$ have the same parity (see Table~\ref{table1}), the corresponding equations for $c_{mn}$ are homogeneous. The only admissible solution is therefore
\begin{equation}
c_{mn} = 0, \quad \text{for } m,n \text{ of the same parity.}
\label{eq56}
\end{equation}

Consider now  $c_{mn}$, where $m, n$ have different parity. From Eq.~(\ref{eq55}), we note that $c_{mn}$ with odd $m$ and even $n$ couple to $c_{mn}$ with even  $m$ and odd $n$. However, if we iterate one more time, then we obtain closed set of equations for either choice of parities:
\begin{widetext}
\begin{align}
&c_{mn} = 
\frac{m \left( m^{4} - 2 m^{2} n^{2} + 9 n^{4} \right)}
{n^{3} (m^{2} - n^{2})^{3} (m^{2} + n^{2})}
+ \frac{16 m n}{\pi^{2} (m^{2} + n^{2})}
\sum_{\substack{m'~\mathrm{odd}\\ n'~\mathrm{even}}}
\frac{
n'^{2} (9 m'^{4} - 2 m'^{2} n'^{2} + n'^{4})
}{
m'^{2} (m^{2} - m'^{2})(n^{2} - n'^{2})
(m'^{2} - n'^{2})^{3} (m'^{2} + n'^{2})
}
\nonumber\\
&
+ \frac{256 m n}{\pi^{4} (m^{2} + n^{2})}
\sum_{\substack{m'~\mathrm{odd}\\ n'~\mathrm{even}}}
\sum_{\substack{m''~\mathrm{even}\\ n''~\mathrm{odd}}}
\frac{
c_{m'' n''} \, m'' n'' m'^2 n'^2
}{
(m'^{2} + n'^{2})
(m^{2} - m'^{2})(n^{2} - n'^{2})
(m'^{2} - m''^{2})(n'^{2} - n''^{2})
} ~\text{for } m~\mathrm{even},~n~\mathrm{odd},
\label{eq57}
\end{align}
\begin{align}
&c_{mn} = 
\frac{n \left( 9 m^{4} - 2 m^{2} n^{2} + n^{4} \right)}
{m^{3} (m^{2} - n^{2})^{3} (m^{2} + n^{2})}
+ \frac{16 m n}{\pi^{2} (m^{2} + n^{2})}
\sum_{\substack{m'~\mathrm{even}\\ n'~\mathrm{odd}}}
\frac{
m'^{2} ( m'^{4} - 2 m'^{2} n'^{2} + 9 n'^{4})}{
(m^{2} - m'^{2})(n^{2} - n'^{2})
n'^{2} (m'^{2} - n'^{2})^{3} (m'^{2} + n'^{2})
}
\nonumber\\
&
+ \frac{256 m n}{\pi^{4} (m^{2} + n^{2})}
\sum_{\substack{m'~\mathrm{even}\\ n'~\mathrm{odd}}}
\sum_{\substack{m''~\mathrm{odd}\\ n''~\mathrm{even}}}
\frac{
c_{m'' n''} \, m'' n'' m'^2 n'^2
}{
(m'^{2} + n'^{2})
(m^{2} - m'^{2})(n^{2} - n'^{2})
(m'^{2} - m''^{2})(n'^{2} - n''^{2})
}~\text{for } m~\mathrm{odd},~n~\mathrm{even}
\label{eq58}
\end{align}
\end{widetext}

These equations cannot be solved in closed form. Rather, we can solve them numerically in an iterative manner. In the first iteration, we drop the third term on the right hand side of Eqs.~(\ref{eq57}) and ~(\ref{eq58}). In the second iteration, we replace $c_{mn}$ appearing in the right hand side with the solution in the first iteration, and so on, to obtain a solution that converges to the right answer. 

To obtain the second moment, we substitute the solution of $c_{mn}$ into  Eq.~(\ref{eq49}), integrate twice, and apply the boundary conditions $E(0)=E(L)=0$. In the limit of large $L$, we obtain
\begin{equation}
\frac{E_P(x)}{J_0^2 L^5}
= \frac{16}{\pi^7} 
\sum_{i,j}  j c_{ij}
\!\!\left[
\frac{\sin\!\left( \frac{(i-j)\pi x}{L} \right)}{(i-j)^2}
+ 
\frac{\sin\!\left( \frac{(i+j)\pi x}{L} \right)}{(i+j)^2}
\right].
\label{eq59}
\end{equation}
The analytical result is compared with Monte Carlo simulations for the second moment of density when $c_{mn}$ are determined upto the third iteration. With each iteration, the analytical result converges to the simulation data.
\begin{figure}
\includegraphics[width=\columnwidth]{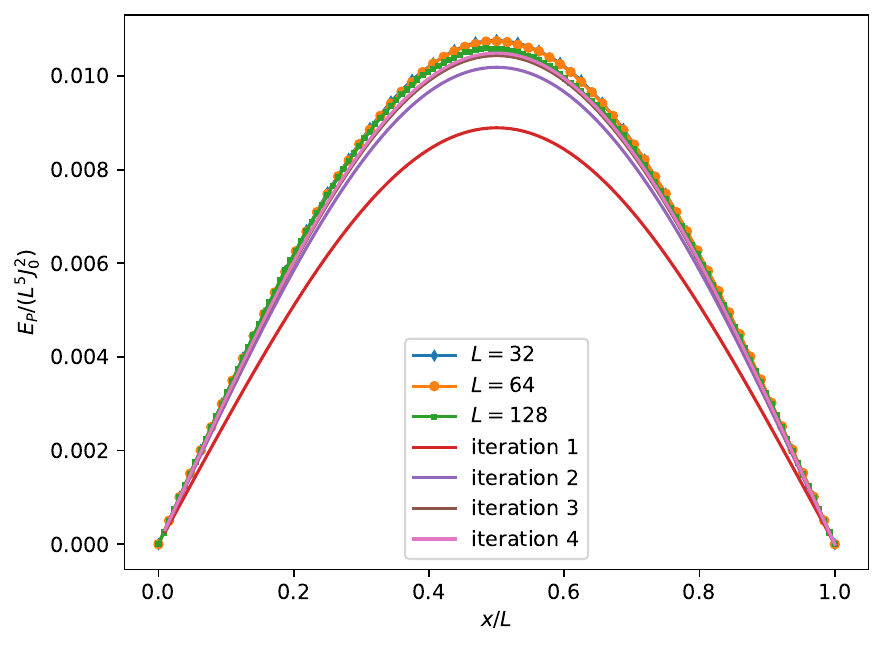}
\caption{The scaled second moment of density, $E_P(x)$, obtained from Monte Carlo simulations is compared with the exact result, Eq.~(\ref{eq59}), when  the $c_{ij}$s are iterated upto one, two and three iterations as described in the text. The data are for periodic boundary conditions.
}
\label{fig:pbc_E}
\end{figure}

\subsubsection{\label{sec:exponents-pbc}Site-dependent exponents}

The determination of the mean density [Eq.~(\ref{eq43})] and second moment [Eq.~(\ref{eq59})] allows us to determine the exponents for the single site mass distribution at the edge (close to condensate) and the bulk. Like for the open boundary conditions, we define the  bulk region to be $L, x \to \infty$ while maintaining $x/L$ constant and  the edge region to be  $L\to\infty$ with $x$ fixed.

\paragraph{\label{sec:bulk-pbc}Bulk scaling:}

\begin{figure}
\includegraphics[width=\columnwidth]{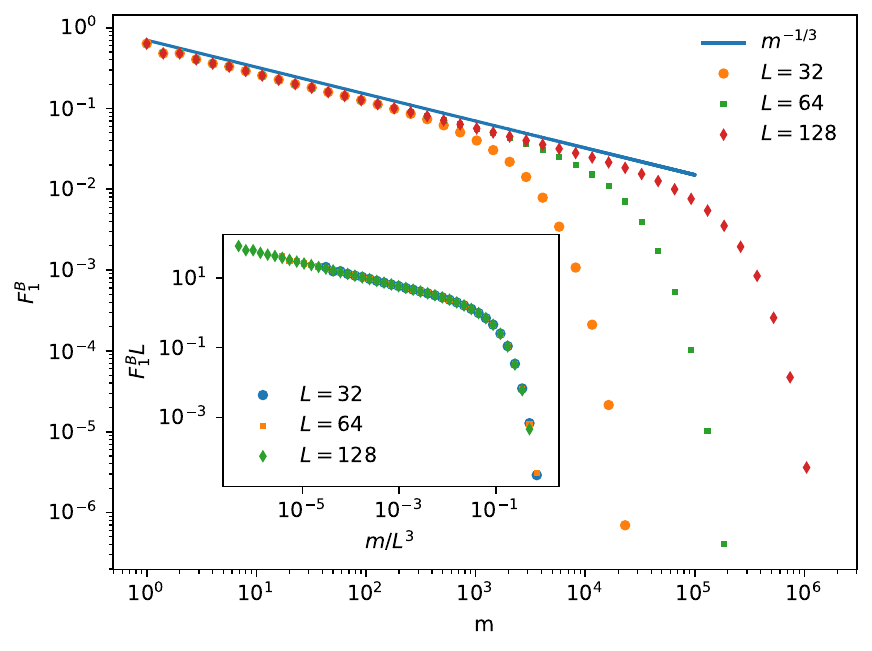}
\caption{The CCDF for the single site mass distribution, $F_1^B$ at the middle site $L/2$ for different system sizes. The data are consistent with the power law $m^{-1/3}$. Inset: The data for different $L$ collapse onto one curve when scaled with the exponents in Eq.~(\ref{eq60}). The data are for periodic boundary conditions.}
\label{fig:bulk-F1-pbc}
\end{figure}

The arguments being very similar to the open boundary conditions, we will be brief. In the bulk $\langle m^2 \rangle \sim L^5$.  This combined with $\langle m \rangle \sim L^2$ from the density profile [see Eq.~(\ref{eq43})],,  yields $\phi(2-\tau_1)=2$ and $\phi(3-\tau_1)=5$, determining $\tau_1$ and $\phi$ in the bulk to be:
\begin{equation}
\tau_{1, pbc}^{B}=\frac{4}{3},  \ \ \ \ \phi_{pbc}^{B}=3.
\label{eq60}
\end{equation}
The result for $\tau_1$ matches  the known answer for $\tau_1$ for an infinite system. Not surprisingly, it also coincides with the result for the open boundary conditions, as we expect the bulk behavior to be independent of the boundary condition. The above results are confirmed using   Monte Carlo simulations, as shown in Fig.~\ref{fig:bulk-F1-pbc}.  The data for the CCDF of the middle site $L/2$ for different system sizes collapse onto one curve when scaled with the exponents given in Eq.~(\ref{eq60}), with the scaling function having a power-law exponent $\tau_1-1=1/3$.

We now consider the multipoint probability distributions as defined in Eqs.~(\ref{eq2})--(\ref{eq4}). As for the single site distribution, we expect the bulk scaling for the multipoint probability distributions $F_n$ to be the same as for the infinite system with $\zeta^B_n$ given by Eq.~(\ref{eq34}), same as that for open boundary conditions.

We measure numerically $F_2$, $F_3$, and $F_4$ for the sites centered around the bulk middle site $L/2$. The data for different $L$ collapse onto a single curve when scaled with $\zeta^B_n$ as in Eq.~(\ref{eq34}) for $n=2, 3, 4$ (see Fig.~\ref{fig:bulk-multi}). We conclude that the scaling for the infinite system continues to hold for multipoint probability distributions in the bulk.
\begin{figure}
\includegraphics[width=\columnwidth]{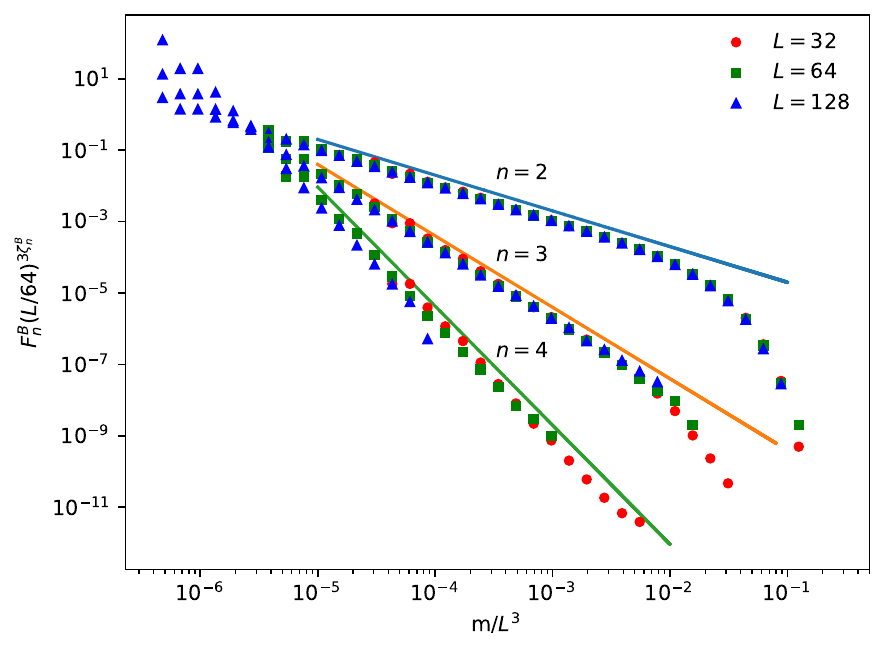}
\caption{The CCDF for multipoint probabilities, $F_n^B$, at the bulk middle site $L/2$ for different system sizes collapse onto one curve when scaled with the exponents in Eq.~(\ref{eq34}) for $n=2, 3, 4$. The solid lines correspond to power-laws with exponents $\zeta_n^B$. The data are for periodic boundary conditions.}
\label{fig:bulk-multi-pbc}
\end{figure}

\paragraph{\label{sec:edge-pbc}Edge scaling:}

While the bulk scaling is independent of the boundary conditions, the edge scaling under periodic boundary conditions could differ from open boundary conditions. This is due to the additional stochastic move of the condensate hopping. We will show that the single site distribution behaves identically while there are differences in the multipoint distributions.

\begin{figure}
\includegraphics[width=\columnwidth]{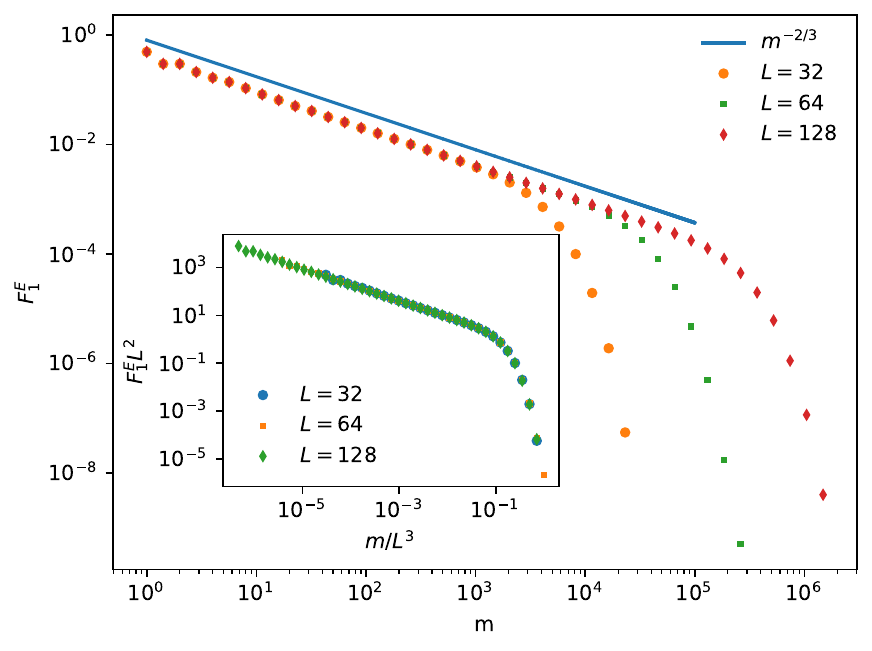}
\caption{The CCDF for the single site mass distribution, $F_1^E$ at the edge site $1$ for different system sizes. The data are consistent with the power law $m^{-2/3}$. Inset: The data for different $L$ collapse onto one curve when scaled with the exponents in Eq.~(\ref{eq61}). The data are for periodic boundary conditions.}
\label{fig:edge-F1-pbc}
\end{figure}

For the single site distribution, the argument is identical to that for the open boundary conditions: $\langle m^2 \rangle \sim L^4$, and $\langle m \rangle \sim L$. Matching moments yields $\phi(2-\tau_1)=1$ and $\phi(3-\tau_1)=4$, determining $\tau_1$ and $\phi$ in the edge to be:
\begin{equation}
\tau_{1, pbc}^{E}=\frac{5}{3},  \ \ \ \ \phi_{pbc}^{E}=3.
\label{eq61}
\end{equation}
We confirm the correctness of the result with Monte Carlo simulations, as shown in Fig.~\ref{fig:edge-F1-pbc}. For the mass distribution of edge site $1$, the data for different system sizes collapse onto one curve when scaled with the exponents given in Eq.~(\ref{eq61}), with the scaling function having a power-law exponent $\tau_1-1=2/3$. 

We now examine the multipoint probability distributions for sites near the edge, as defined in Eqs.~(\ref{eq2})--(\ref{eq4}). It is tempting to argue that the scaling of the multipoint correlations will be identical to those for the open boundary conditions since in one dimension, both boundary conditions correspond to one special lattice point. However, for periodic boundary conditions, the condensate hops, which in its frame of reference corresponds to all the other particles diffusing in a synchronized matter. It is not apriori clear whether this is relevant for scaling. 

Consider the time evolution of $F_k(m, x)$, the CCDF for the order $k$ multipoint distribution at $x$. The hopping of the condensate introduces a diffusion term in BBGKY hierarchy:  $d F_k(m,x)/dt \sim D \nabla^2 F_k(m,x)$. For this diffusion term to be relevant, it must be non-vanishing, yet well behaved at the edges. Consider the first non-trivial correlation $k=2$. For the term to be relevant  $F_k(m,x) \sim x^2$ for small $x$. Assuming this dependence allows us to determine the scaling exponent of $F_2(m,x)$ as follows. $F_2(m,x)$ will have the scaling form
\be
F_2(m,x) = \frac{1}{x^{\phi \zeta_2^B}} h_2\left(\frac{m}{x^\phi} \right).
\label{eq62}
\ee
In the bulk, i.e. when the argument of $h_2$ is small, we expect no dependence on $x$, implying that $h(y)\sim y^{-\zeta_2^B}$ for $y\ll 1$. When $y\gg 1$, corresponding to edge sites, we expect $h(y)\sim y^{-\zeta_2^B-2/\phi}$, so that the dependence of $F_2(m,x)$ on $x$ is quadratic. This implies that
\be
\zeta_{2,pbc}^E= \zeta_2^B+\frac{2}{\phi}.
\label{eq63}
\ee
Substituting $\zeta_2^B=1$ and $\phi=3$, we obtain $\zeta_{2,pbc}^E=5/3$ which is different from $\zeta_{2, obc}^E=2$. 

\begin{figure}
\includegraphics[width=\columnwidth]{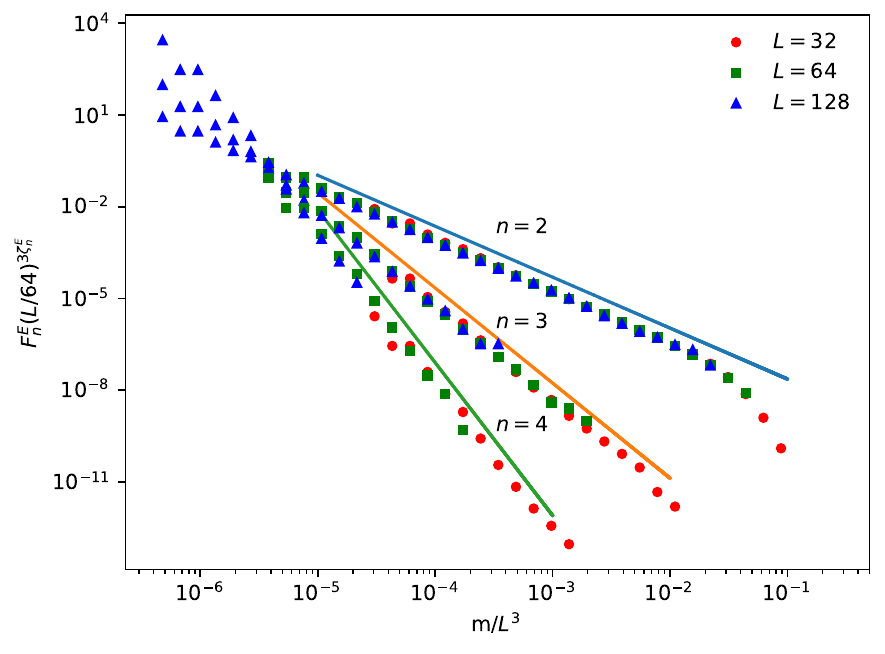}
\caption{The CCDFs of the multipoint probabilities $F_n^E$ near the edge site ($x=1$) for different system sizes. The data collapse onto a single curve when scaled using $\zeta_{2, pbc}^E=5/3$,  $\zeta_{3, pbc}^E \approx 28/9$, and $\zeta_{4, pbc}^E \approx 5$. The solid lines indicate power laws with exponents $\zeta_{n,pbc}^E$. The data are for periodic boundary conditions. }
\label{fig:edge-multi-pbc}
\end{figure}

This argument is based on the relevance of the diffusion term due to the hopping of the condensate, and only simulations can provide the actual test. We numerically evaluate $F_2$, $F_3$, and $F_4$ for sites near the edge ($x=1$). As shown in Fig.~\ref{fig:edge-multi-pbc}, data for different system sizes for $F_2$ collapse onto a single curve when scaled with $\zeta_{2, pbc}^E=5/3$. For higher order probabilities, we measure the scaling exponents numerically. We numerically find that  $\zeta_{3, pbc}^E\approx 3.11 \approx 28/9$ and $\zeta_{4, pbc}^E \approx 5.0$. For these choices, we obtain good data collapse (see Fig.~\ref{fig:edge-multi-pbc}). The values of these exponents are well-described by the equation
\be
\zeta_{n, pbc}^E=\frac{2n^2 + 3 n+ 1}{9},~~n=1,2, \ldots,
\label{eq64}
\ee
which reproduces the results for $n=1,2$.

\section{Mass dependent diffusion \label{sec:mass-dependent}}

Building on our understanding of finite-size effects in the Takayasu model, we now generalize the arguments to a system with mass-dependent diffusion, where the diffusion constant of a particle with mass \( m \) scales as \( m^{-\alpha} \). Realistic models of physical and biological systems often involve such mass-dependent transport. For instance, according to the Stokes–Einstein relation for droplets, the diffusion constant is inversely proportional to the droplet radius, or equivalently, to \( m^{1/d} \) in $d$ dimensions. In ballistic aggregation, the particle velocity scales as \( m^{-1/2} \). The inclusion of mass-dependent diffusion can substantially influence condensate formation. For example, in models with chipping, slow diffusion has been shown to inhibit condensate formation~\cite{rajesh2002aggregate}.

We first derive the power-law exponents in the limit of an infinite system size. Since exact analytical results are no longer attainable, we employ scaling arguments based on the concept of a catchment area~\cite{rajesh2004nonequilibrium}. Let \( M_t \) denote the typical large mass at time \( t \), and \( L_t \) the region it explores during this time. In the presence of input, it follows that \( M_t \sim t L_t \). To determine the time evolution of \( L_t \), we incorporate mass-dependent diffusion:
\be
\frac{d L_t^2}{dt} \sim D(M_t) \sim M_t^{-\alpha} \sim (t L_t)^{-\alpha}.
\label{eq65}
\ee
Solving for $L_t$  gives
\be
L_t \sim t^{\frac{1-\alpha}{2+\alpha}}.
\label{eq66}
\ee
This immediately implies \( M_t \sim t^\delta \), where \( \delta = 3/(2+\alpha) \). Since the total mass increases linearly with time, i.e., \( \delta (2-\tau) = 1 \), we obtain
\be
\tau_1^B = \frac{4 - \alpha}{3}, \quad \alpha < 1.
\label{eq67}
\ee

For finite systems, we determine the scaling of the largest mass with system size \( L \) as follows. The characteristic time \( t_c \) required to reach the steady state is obtained by setting \( L_t = L \), yielding \( t_c \sim L^{(2+\alpha)/(1-\alpha)} \). During this time, the mean mass per site is proportional to  \( J t_c \). Equating this to the scaling form \( L^{\phi (2 - \tau)} \), we find
\be
\phi = \frac{3}{1 - \alpha}.
\label{eq68}
\ee

To examine spatial variations near the edges, we consider the site-dependent single-site mass distribution \( P_1(m,x) \), which has the scaling form
\be
P_1(m,x) \sim m^{-\tau_1^B} f\!\left(\frac{m}{x^\phi}\right),
\label{eq69}
\ee
where the scaling function \( f(y) \) approaches a constant for \( y \ll 1 \). For \( y \gg 1 \), we argue that \( f(y) \sim y^{-1/\phi} \), ensuring that the distribution varies linearly with \( x \) near the edges. This linear dependence is required for the mass current, proportional to \( dP(m,x)/dx \), to remain constant. Consequently,
\be
\tau_1^E = \tau_1^B + \frac{1}{\phi} = \frac{5 - 2\alpha}{3}.
\label{eq70}
\ee

For \( \alpha = 0 \), these results recover the exact expressions obtained earlier. For \( 0 \leq \alpha < 1 \), the exponents remain unchanged for both open and periodic boundary conditions. When \( \alpha < 0 \), massive particles diffuse faster than lighter ones. Under open boundary conditions, the scaling arguments remain valid for \( \alpha > -2 \), for which \( \tau_b < 2 \). In contrast, for periodic boundary conditions, the steady state becomes trivial: the condensate rapidly absorbs all incoming particles, leaving a single diffusing condensate in steady state. 

\begin{figure}
\includegraphics[width=\columnwidth]{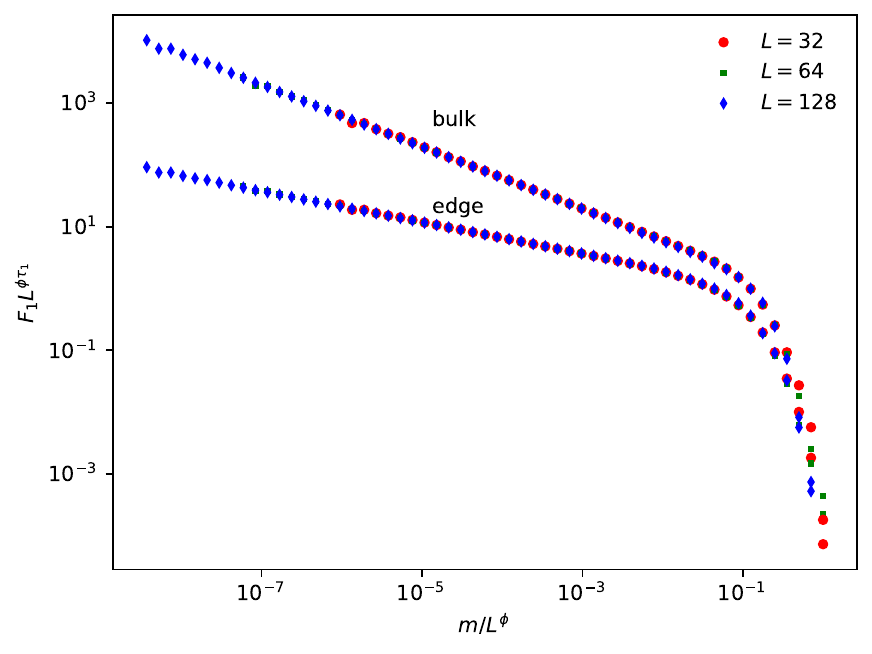}
\caption{The CCDF for the  single-site mass distribution \( F_1 \) at the middle site (\( L/2 \)) and edge site (\( 1 \)) for different system sizes in the case of mass-dependent diffusion with \( \alpha = 1/4 \), i.e., $D(m) \sim m^{-1/4}$. The data for different \( L \) collapse onto a single curve when scaled with the exponents in Eqs.~(\ref{eq67}), (\ref{eq68}), and (\ref{eq70}). All data correspond to open boundary conditions.}
\label{fig:massdependent-F1}
\end{figure}

To validate the scaling exponents, we performed simulations with open boundary conditions. We verified that the results remain unchanged under periodic boundaries for \( \alpha > 0 \). Figure~\ref{fig:massdependent-F1} shows data for the single-site complementary cumulative distribution function (CCDF) \( F_1 \) at the middle site (\( L/2 \)) and edge site (\( 1 \)) for \( \alpha = 1/4 \). For this case, the predicted exponents are \( \tau_1^B = 5/4 \), \( \tau_1^E = 3/2 \), and \( \phi = 4 \). Data for different system sizes collapse onto a single curve for both edge and bulk sites when scaled with these exponents.

\section{\label{sec:conclusions} Summary and conclusions}

We have presented exact analytical and numerical results for the Takayasu model under both open and periodic boundaries, deriving closed-form expressions for the steady-state density, correlations, and mean-squared mass.
These results demonstrate a clear crossover in the single-site mass distribution—from the bulk power law $P(m) \sim m^{-4/3}$ to an edge power law $P(m) \sim m^{-5/3}$. In periodic systems, this ``edge'' corresponds to sites near the condensate, while in open systems it occurs near the physical boundaries. Surprisingly, the equivalence between the two boundary conditions break down in the case of multipoint distributions. In particular, we showed that the homogeneity exponents for the multipoint distributions near the edge are different for periodic and open boundary conditions. Finally, we generalized the results to the case of mass-dependent diffusion.

The exact solution provides a sharp criterion to distinguish between bulk and edge regimes. The edge limit corresponds to $L \to \infty$ with $x$ fixed, while the bulk limit corresponds to $L, x \to \infty$ with $x/L$ fixed. These findings clarify earlier numerical results~\cite{negi2024condensate}, which identified a boundary layer containing a finite fraction of the lattice sites (of extent $x/L \lesssim 0.1$) where edge effects dominate.

The exponent $\tau_1 = 5/3$ appears in several seemingly unrelated contexts involving infinite systems. For instance, in the charge model with injection of oppositely charged monomers, the steady-state charge distribution follows $P(q) \sim q^{-5/3}$~\cite{sire1995correlations}. A similar scaling arises in the two-point joint distribution $P(m_1,m_2)$ of the Takayasu model, where numerical evidence suggests that $P(m_1,m_2) \sim m_1^{-5/3}$ for $m_1 \ll m_2$~\cite{connaughton2008constant}. The edge site adjacent to a large condensate plays an analogous role to the small-mass neighbor of a large cluster, linking the finite-system edge behavior to the two-point scaling in infinite systems. From this perspective, the exponent $5/3$ derived here provides an exact characterization of the $m_1\ll m_2$ asymptotic regime of $P(m_1,m_2)$ for the infinite system. 

It is natural to consider a system with a condensate  to be equivalent to the system with open boundary conditions in one dimensions, as both systems have an effective sink site where mass disappears. This would explain why both systems exhibit the same edge exponent $\tau_1 = 5/3$. However, contrary to the naive expectation that a condensate acts simply as a sink, we find that its stochastic motion introduces a long-range correlation that fundamentally alters the scaling of multipoint distributions near the edge.  We trace this difference between the boundary conditions to the hopping of the condensate, which effectively translates to a co-ordinated hopping of all the other particles, introducing an extra diffusion term in the BBGKY hierarchy. By considering the relevance of such diffusive terms, we were able to determine the scaling of the two-point probability distribution using scaling arguments.

In higher dimensions, finite-size effects are expected to depend sensitively on boundary conditions. For periodic systems, sites neighboring the condensate should behave similarly to the two-point distribution in the infinite system. Since $d=2$ is the upper critical dimension, both bulk and edge sites are expected to exhibit $\tau_1 = 3/2$ with logarithmic corrections. For open systems, however, different types of boundaries should lead to distinct scaling: we conjecture $\tau_1 = 7/4$ for edge sites and $\tau_1 = 2$ for corner sites. Extending the present framework to verify these predictions remains an important direction for future work.

The introduction of finite boundaries also induces spatial mass currents—toward the edges for open systems and toward the condensate for periodic systems—in addition to the mass-space flux generated by aggregation. The latter gives rise to the constant-flux relation, under which the two-point distribution $P(m_1,m_2)$ for neighboring sites has a universal scaling dimension of three in all dimensions~\cite{connaughton2007constant,connaughton2008constant}. While this relation holds in the bulk, it breaks down near the edges, where the dominant spatial currents are of order $\mathcal{O}(L)$ compared to the $\mathcal{O}(1)$ mass-space flux, providing a consistent physical interpretation of the edge anomaly.

Similar finite-size or boundary-induced effects are expected in other mass-conserving models such as the conserved mass aggregation model~\cite{majumdar1998nonequilibrium,rajesh2001exact,sachdeva2013condensation} and the in–out model~\cite{majumdar2000phase,rajesh2004nonequilibrium}. On the experimental front, aggregation processes in biomolecular condensates and cellular phase separation~\cite{brangwynne2011active,lee2023size,arsenadze2024anomalous,liu2021scale,leggett2019motility} are essentially in a finite geometry, and offer promising avenues to observe the edge-induced scaling and crossover phenomena predicted here.

\section{Acknowledgments}
RBR thanks The Institute of Mathematical Sciences, Chennai for a summer student fellowship.


%

\end{document}